\documentclass[aps,prb,twocolumn,showpacs]{revtex4}
\usepackage{graphicx}
\usepackage{epstopdf}
\begin{document}
\title{Pressure-temperature phase diagram of ferromagnetic superconductors}

\author{V. P. Mineev}

\affiliation{Commissariat \`a l'Energie Atomique, INAC / SPSMS, 38054
Grenoble, France}
\date{\today}

\begin{abstract}
The symmetry approach to the description of the $(P,T)$ phase diagram of ferromagnet superconductors with triplet pairing is developed. Taking into account the recent experimental observations made on $UCoGe$ it is considered the case of  a crystal with orthorhombic structure and strong spin-orbital coupling. It is shown that formation of ferromagnet superconducting state from a superconducting state is inevitably accompanied by the first order type transition.
\end{abstract}


\maketitle

There are several metallic compounds demonstrating the coexistence of superconductivity and
itinerant ferromagnetism. These are $UGe_2$, \cite{Saxena} $URhGe$, \cite{Aoki} and recently revealed $UCoGe$.\cite{Huy07}
The large band splitting and the high low temperature value of upper critical field \cite{Huxley01,Hardy051,Huy08} in uranium ferromagnetic superconductors indicate that here we deal with 
Cooper pairing in the triplet state.
The superconductivity in $UGe_2$ and $URhGe$ arises at temperatures far below the corresponding Curie temperature and coexists with ferromagnetism in some pressure interval such that at the $(P,T)$ phase diagram the whole   region occupied by the superconducting state is situated inside of more vast ferromagnetic region. \cite{Huxley03,Hardy052} The phase diagram of  new 
ferromagnetic superconductor $UCoGe$  is found to be qualitatively different.
\cite{Hassinger} At ambient pressure the ferromagnetism ($T_{Curie}=2.8 K$) coexists with superconducting state  ($T_{sc}=0.8 K$)
arising close to ferromagnetic instability. Then at applied pressure  the Curie temperature is diminished 
such that no indication of ferromagnetic order has been observed above $P^*\approx 1~GPa$. The resistive superconducting transition is, however, quite stable in temperature and persists up to the highest measured pressure of about $2.4~ GPa$.

The apparent intersection of pressure dependent transition lines $T_{Curie}(P)$ and $T_{sc}(P)$
raises the problem about the general properties of  $(P,T)$ phase diagram including the region  of coexistence of  ferromagnetic and triplet pairing superconducting states. 
It has its own peculiarity   other than that of singlet superconductivity  coexisting
with ferromagnetism in form is known
as the Anderson-Suhl or cryptoferromagnetic superconducting state (for review see \cite{Buzdin}) characterized by formation of a transverse domain-like
magnetic  structure. The structure period or the domain size is larger than interatomic distance and smaller than the superconducting coherence length that decreases the depairing effect of the exchange field.
The latter is irrelevant in the case of triplet superconductivity, hence there is no reason for formation of a cryptomagnetic state. 

The region of coexistence of superconductivity and ferromagnetism is separated from the normal state by the region of ferromagnet normal state at $P<P^*$ and by the region of superconducting state at $P>P^*$. 
Let us look first at $P<P^*$ region.

All the mentioned ferromagnetic superconductors are the metals with orthorhombic symmetry and magnetic moment directed along of one of the crystallographic axis which we choose  as $\hat z$ axis.
The symmetry of the normal paramagnetic state is determined by the elements of group
\begin{equation}
G=D_2\times U(1)\times R,
\label{e-1}
\end{equation}
where $D_{2}=(E, C_{2}^{z},
C_{2}^{x}, C_{2}^{y})$ is the point symmetry group, $U(1)$ is the group of gauge transformations and 
$R$ is the time reversal operation. At the transition to the normal ferromagnet state the  symmetry reduces to the 
\begin{equation}
G_M=D_2^M(C_2^z)\times U(1),
\label{e0}
\end{equation}
where $D_2^M(C_2^z)=(E,C_2^z,RC_2^x,RC_2^y)$ is so called magnetic class \cite{Landau}.
The symmetries and the order parameters of unconventional superconducting states
arising from the normal state with ferromagnetic order in orthorhombic crystals with strong spin-orbital coupling have been pointed out in the paper.
\cite{Mineev} They belong to two different corepresentations $A$ and $B$ such that two ferromagnetic superconducting classes of different symmetry denoted $A_1, A_2$ and $B_1, B_2$ are related to each of them correspondingly. Then  it was pointed out \cite{Sam,Cham} that superconducting states in triplet ferromagnet superconductors represent a special type of two band superconducting states, namely, consisting of pairing states occupied either by 
 spin up or spin down electrons. The two-component order parameters have a  form
\begin{equation}
{\bf d}({\bf k})=\frac{1}{2}
[-(\hat{x}+i\hat{y})\Delta_{\uparrow}¥({\bf k})+
(\hat{x}-i\hat{y})\Delta_{\downarrow}¥({\bf k})].
\label{e1}
\end{equation}
Here $\hat{x}$, $\hat{y}$  are the
unit vectors of the spin 
coordinate system pinned to the crystal axes.
\begin{equation} 
\Delta_{\uparrow}¥({\bf k})=-\eta_{\uparrow}¥f_{-}¥({\bf
k}),~~~~ \Delta_{\downarrow}¥({\bf k})=\eta_{\downarrow}¥
f_{+}¥({\bf k}).
\label{e2}
\end{equation}
Functions $f_{\pm}¥({\bf k})$
are
odd functions of momentum directions of pairing particles on the Fermi
surface.   The general forms of these functions for the
different co-representations in ferromagnetic superconductors with
orthorhombic  symmetry are  listed in the papers. \cite{Mineev,Cham}

The complex order parameter amplitudes
$\eta_{\uparrow}$ and $\eta_{\downarrow}¥$ 
are not completely independent:
\begin{equation}
\eta_{\uparrow}¥=|\eta_{\uparrow}¥|e^{i\varphi}¥,
~~~\eta_{\downarrow}¥=\pm|\eta_{\downarrow}¥| e^{i{\varphi
}¥}.
\label{e3}
\end{equation}
Thus, being different by their modulos they have the same phase with an
accuracy $\pm \pi$. At the phase transition to the normal ferromagnet state the amplitudes of the both components of the superconducting order parameter  turn to zero.

The superconducting order parameter form (\ref{e1}) is quite natural in ferromagnetic state with
spin up - spin down band splitting. We shall use another form of the order parameter which is 
more appropriate when
 the ferromagnetism (but not superconductivity !) is suppressed by pressure, and the band splitting plays
no role. So, we choose the order parameter consisting of  sum of real and imaginary parts. For the states $S_{A_1}$ and $S_{A_2}$ they are
 \begin{equation}
{\bf d}_{A_1}({\bf k})=\eta_1\mbox{\boldmath$\varphi$}_1({\bf  k})+
i\eta_2\mbox{\boldmath$\varphi$}_2({\bf k}),
\label{e4}
\end{equation}
\begin{equation}
{\bf d}_{A_2}{\bf k})=i\eta_1\mbox{\boldmath$\varphi$}_1({\bf  k})+
\eta_2\mbox{\boldmath$\varphi$}_2({\bf k}),
\label{e5}
\end{equation}

where
 \begin{equation}
\mbox{\boldmath$\varphi$}_1({\bf  k})=u_1\hat x k_x+u_3\hat y k_y,~~~~
\mbox{\boldmath$\varphi$}_2({\bf k})=u_2\hat x k_y+u_4\hat y k_x,
\label{e6}
\end{equation}
and $u_1,....$ are real functions of $k_x^2, k_y^2, k_z^2$.
Due to the property (\ref{e3}) the amplitudes $\eta_1$ and 
$\eta_2$ can be chosen having equal phase factors.
Both states with the order parameters  (\ref{e4}) and (\ref{e5}) belong to the same corepresentation $A$ 
and have common critical temperature of transition from the normal ferromagnet state. But they obey
the different symmetry (belong to different ferromagnetic superconducting classes)
\begin{equation}
G_{A_1}=D_{2}(C_{2}^{z})=
(E, C_{2}^{z}, RC_{2}^{x}, RC_{2}^{y}),
\label{e7}
\end{equation}
and
\begin{equation}
G_{A_2}=\tilde D_{2}(C_{2}^{z})= (E, C_{2}^{z},
RC_{2}^{x}e^{i\pi},RC_{2}^{y}e^{i\pi}).
\label{e8}
\end{equation}
They also differ each other by the direction of the Cooper pair spin momentum
\begin{equation}
 {\bf S}_{A_1}=i\frac{\langle{\bf d}_{A_1}^*\times{\bf d}_{A_1}\rangle}{\langle{\bf d}_{A_1}^*{\bf d}_{A_1}\rangle}
\label{e9}
\end{equation}
\begin{equation}
{\bf S}_{A_2}=-{\bf S}_{A_1}.
\label{e10}
\end{equation}
Here the angular brackets mean the averaging over the Fermi surface.

At the second order type transition from the normal  ferromagnet to the superconducting ferromagnet the states $S_{A_1}$ and $S_{A_2}$  appear in the ferromagnet domains with the opposite direction of magnetization \cite{Mineev}. Thus the order parameter
of superconducting state proves to be coordinate dependent and 
can be written as 
 \begin{equation}
{\bf d}({\bf r},{\bf k})=e^{i\alpha({\bf r})}\eta_1({\bf r})\mbox{\boldmath$\varphi$}_1({\bf  k})+
e^{-i\alpha({\bf r})+i\pi/2}\eta_2({\bf r})\mbox{\boldmath$\varphi$}_2({\bf k}),
\label{e11}
\end{equation}
where the phase $\alpha =0$ in the magnetization up domains with $S_{A_1}$  superconducting state
and $\alpha=\pi/2$ in the magnetization down domains with $S_{A_2}$  superconducting state. The phase is changed near   atomic thickness ferromagnet domain walls in the layer of the order of supeconducting coherence length. At the phase transition to the normal ferromagnet state the amplitudes of the both components of the superconducting order parameter (\ref{e11}) turn to zero.

Let us consider now what is going on at $P>P^*$. Here, at temperature decrease, the system can pass 
from the normal to the superconducting state. It is quite natural to think that it happens by means of the second order type transition to conventional superconducting state $S_1$ with 
the order parameter
 \begin{equation}
{\bf d}_{S_1}({\bf k})=\eta_1\mbox{\boldmath$\varphi$}_1({\bf  k})
\label{e12}
\end{equation}
having the full point symmetry of the crystal lattice and relating to the trivial superconducting class 
\begin{equation}
G_{s1}=D_2\times R.
\label{e13}
\end{equation}
As any one component  superconducting state formed directly from a nonmagnetic normal state by means of the second order type transition this state
is nonmagnetic. \cite{MinSam} The spontaneous magnetism appears at  a transition to the ferromagnet state where the superconducting order parameter acquires 
the second component.  
By means of the second order phase transition the state $S_1$ can turn  only to the homogeneous superconducting ferromagnet state $S_{A_1}$ with symmetry group $G_{A_1}$ which is the subgroup
of $G_{s1}$.
So, we come to the ferromagnet state with only one type of superconducting domains. The second type of superconducting domains formed by the state $S_{A_2}$  can not be smoothly created  from the state $S_1$ simply because its symmetry group $G_{s1}$ does not contain $G_{A_2}$ as its subgroup.

Similar situation occurs if instead of conventional superconducting state
the system pass first to nonconventional 
superconducting state with the order parameter
 \begin{equation}
{\bf d}_{S_2}({\bf k})=\eta_2\mbox{\boldmath$\varphi$}_2({\bf  k}),
\label{e14}
\end{equation}
having the symmetry group or superconducting class 
\begin{equation}
G_{s2}=D_2^s(C_2^z)\times R,~~~~~D_2^s=(E, C_2^z, e^{i\pi}C_2^x, e^{i\pi}C_2^y).
\label{e15}
\end{equation}
This state  by the second order phase transition can be transformed only to the homogeneous superconducting ferromagnet state $S_{A_2}$ with symmetry group $G_{A_2}$ which is the subgroup
of $G_{s2}$. The  superconducting domains formed by the state $S_{A_1}$  can not be smoothly created  from the state $S_2$ because its symmetry group $G_{s2}$ does not contain $G_{A_1}$ as its subgroup. Thus, the formation of multi domain superconducting ferromagnet state (\ref{e11})
from homogeneous superconducting state either conventional type  with an order parameter (\ref{e12}), or nonconventional type with an order parameter (\ref{e14}) by means the second order type transitions is impossible. 

We come to the conclusion that formation of multi domain ferromagnet superconducting state at $P>P^*$ is inevitably related  with first order type transition.
Either, at temperature decrease first we have second order type transition to the homogeneous superconducting state followed at lower temperature by the first order type transition to the multi domain superconducting ferromagnet state (see Fig.1a). 
\begin{figure}[tbp]
\includegraphics[scale=0.5]{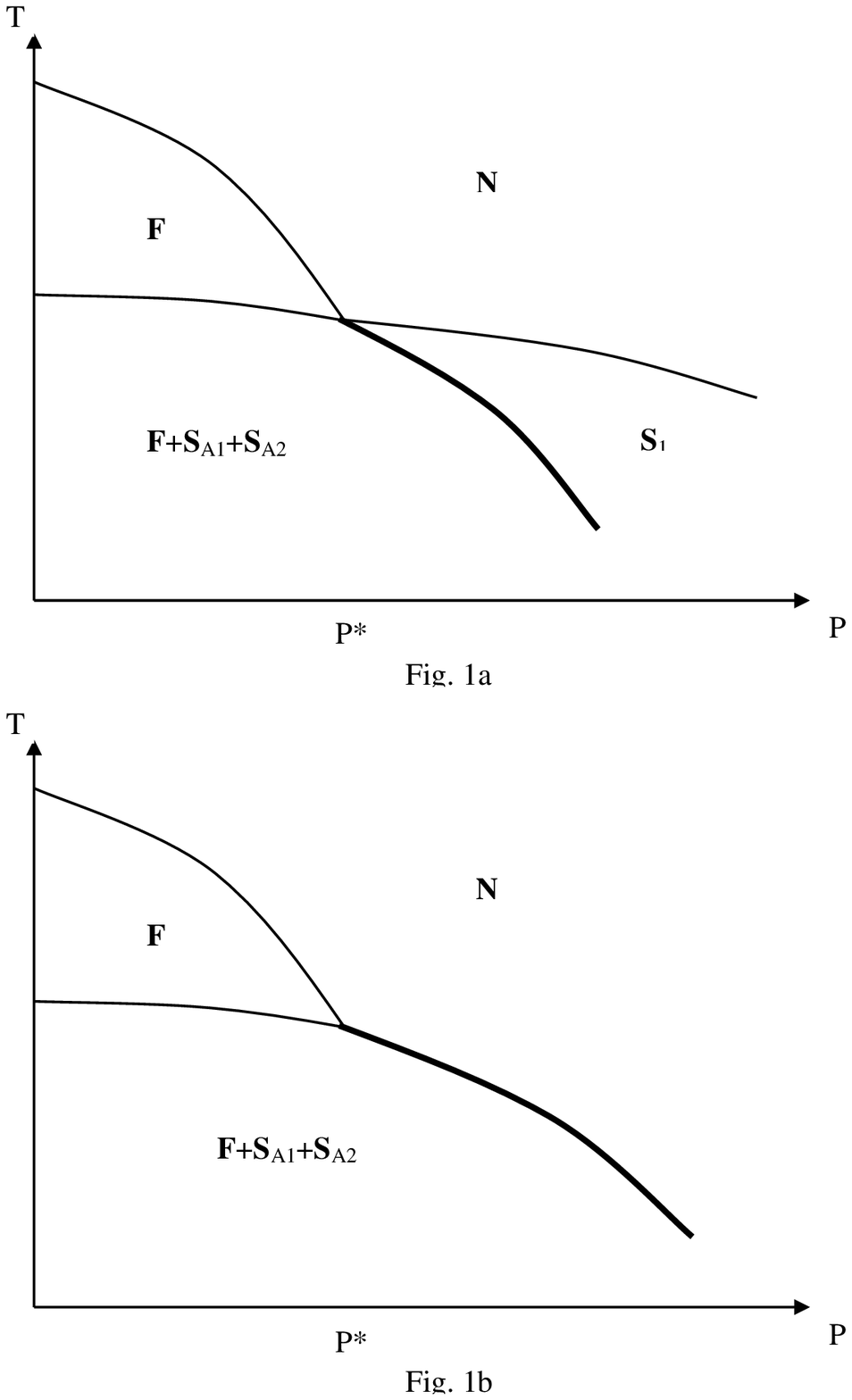}
\caption{Schematic pressure - temperature phase diagrams: (a) and (b) (see the text). Here, $N$ is the normal paramagnetic phase, $F$ is the ferromagnet phase, $S_1$ is the homogeneous 
superconducting phase, $F+S_{A_1}+S_{A_2}$ is multi domain ferromagnet superconducting phase.
The thin lines are the lines of the second order transithions, the thick lines are the lines of the first 
order transitions.}
\label{fig.1}
\end{figure}
Or,  there is only one first order type transition 
from the normal paramagnet state to the  multi domain superconducting ferromagnet state 
(see Fig.1b). 
Both of these possibilities are compatible with 
an important observation made by resistivity measurements on polycrystalline specimens of $UCoGe$
\cite{Hassinger} that the superconducting phase transition in the paramagnetic state  
becomes much broader than in the magnetically ordered phase.  To establish complete pressure-temperature phase diagram of single crystal $UCoGe$ the application of the other experimental methods 
is desirable. 

The external field of about 100 Oe directed along the easy magnetization axis of $UCoGe$ transforms it from the multi domain to the single domain ferromagnet \cite{Huy08}. 
The transformation process  should accompanied by weakening of the first order phase transition depicted at Fig.1 and smooth conversion of it to the second order type transition.

In conclusion, we have found the pressure-temperature phase  diagram of a ferromagnetic triplet superconductor with strong spin-orbital coupling and orthorhombic crystal structure. Its shape  is dictated by the symmetry of the order parameter. The experimental revelation of such type phase diagram
in $UCoGe$ or similar compounds
can serve as the direct indication to the type of superconducting state we deal with.

\section*{ACKNOWLEDGMENTS}

The author is indebted to M. Houzet and  A. Buzdin for valuable discussions, to D.Aoki and J.Flouquet for the interest to the work and to A. Muratov and S. Blundell for the kind assistance in the manuscript preparation.

\end{document}